\documentclass[12pt]{iopart}
\usepackage{iopams}
\usepackage{cite}
\usepackage{graphicx}

\newcommand{\bear}{\begin{eqnarray}}
\newcommand{\eear}{\end{eqnarray}}
\newcommand{\be}{\begin{equation}}
\newcommand{\ee}{\end{equation}}
\newcommand{\beqn}{\begin{eqnarray}}
\newcommand{\eeqn}{\end{eqnarray}}
\newcommand{\beqnn}{\begin{eqnarray*}}
\newcommand{\eeqnn}{\end{eqnarray*}}

\def\vep{\varepsilon}
\def\vf{\varphi}

\begin{document}

\title{Rotating quantum Gaussian packets }

\author{V V Dodonov }


\address{
 Instituto de F\'{\i}sica, Universidade de Bras\'{\i}lia,
Caixa Postal 04455, 70910-900 Bras\'{\i}lia, DF, Brazil }

\begin{abstract}
We study two-dimensional quantum Gaussian packets with a fixed value of mean angular momentum. This value is the sum of two independent
parts: the `external' momentum related to the motion of the packet center and the `internal' momentum due to quantum fluctuations.
The packets minimizing the mean energy of an isotropic oscillator with the fixed mean angular momentum are found. 
They exist for `co-rotating' external and internal motions, and they have nonzero correlation coefficients between coordinates 
and momenta, together with some (moderate) amount of quadrature  squeezing. Variances of angular momentum and energy
are calculated, too. Differences in the behavior of `co-rotating' and `anti-rotating' packets are shown.
The time evolution of rotating Gaussian packets is analyzed, including the cases of a charge in a homogeneous 
magnetic field and a free particle. In the latter case, the effect of initial shrinking of packets with big enough coordinate-momentum
correlation coefficients (followed by the well known expansion) is discovered.
This happens due to a competition of `focusing' and `de-focusing' in the orthogonal directions.

\end{abstract}

Keywords: minimal energy packets, isotropic harmonic oscillator, shrinking/expanding free packets, correlation coefficient, squeezing, 
magnetic field, internal/external rotations competition

\pacs{03.65.-w, 03.75.-b  } 
\ead{vdodonov@fis.unb.br} 

\submitto{\JPA}

\section
{Introduction}
Quantum Gaussian wave packets were considered at the dawn of quantum mechanics by
Schr\"odinger \cite{Schrodpac}, Kennard \cite{Kenn} and Darwin \cite{Darwin}. 
Since that time, different properties of such packets were studied by many authors.
We are interested here in {\em two-dimensional\/} packets. They were discussed in detail, e.g., in original papers,
books and reviews
 \cite{Holevo,Milb84,Schurep,Little86,SiSuMu87,183vol}, where the main emphasis was made on the {\em squeezing\/} properties,
or in more recent papers \cite{Seraf04,DDM05,Rend05,Ades05,Seraf07,Wang07,Weed12}, 
where the problems of quantum information (in particular {\em entanglement properties\/}) were the focus of studies.

The aim of the present paper is to study properties of  Gaussian packets possessing nonzero mean values of the
angular momentum operator. 
Different special cases of such packets were considered by many authors for a long time.
In particular, packets whose centers move along some circles arise naturally in the description of quantum
 charged particles placed in a uniform magnetic field. An example of such packets was given for the first time by
Darwin \cite{Darwin}. Similar `circulating wave packets' were considered later by Husimi \cite{Hus53}. These packets were
generalized and interpreted as {\em coherent states\/} of a charged particle (or an oscillator) in a magnetic field by Malkin and Man'ko 
\cite{MM69,MMT70} and other authors \cite{Tam71,Var84}. 
Using such packets one can simplify substantially the analysis of many physical effects, in particular the Landau diamagnetism \cite{FeKa70}. 
The Gaussian packets in a magnetic field were studied also in \cite{KimWei73,Santos09}.
Two-dimensional coherent states in rotated frames were shown to be useful for the
description of rotational properties of nuclei in the framework of the so called `cranked oscillator' model \cite{GulVol80,GV80}.
More recently, rotated Gaussian packets or Gaussian packets in rotating frames were studied in \cite{Hac96,Wun02,Bracher11,Ross11,Ross14}.
Mathematically equivalent `twisted' paraxial optical beams with nonzero orbital angular momentum were studied, e.g., in 
\cite{Allen92,Simon93,San13,Karimi14}.

During the past two decades, the so called `vortex states' of light beams with nonzero orbital angular momentum received  
much attention \cite{Agar97,Berry04,Padg04,Abram04,Berry08,Gotte08}. 
In parallel, many authors studied vortex states of quantum massive particles  \cite{Arvi97,BB00,Bliokh07,Bliokh12,Gall12,Hac13,Karlov15}. 
Recently such states were created in different experiments \cite{Uchida10,Verb10,McMorr11,Grillo15}.
It was pointed out in the cited papers that matter wave beams with orbital angular momentum can find applications in
different areas, such as condensed-matter spectroscopy, electron microscopy and particle physics. In particular,
they can be used in the study of magnetic properties of materials and for manipulating nanoparticles.
Therefore a study of  properties of rotating packets that were not considered earlier seems to be an interesting and timely task.

The new question answered in this paper is as follows: what is the {\em minimal mean energy\/} of {\em quantum packets\/} with a
 fixed value of {\em mean angular momentum}?
This question has sense, e.g., for a free isotropic oscillator or an isotropic oscillator placed in a homogeneous
magnetic field (with a free particle moving in the magnetic field as a special case).

Solutions to the stationary Schr\"odinger equation for the two-dimensional harmonic oscillator of mass $M$ and frequency $\omega$ 
in the polar coordinates are known since the very beginning of quantum mechanics \cite{Schrod-osc}:
\be
\fl
\psi_{n_r m}(r,\vf) = N r^{|m|}\Phi\left(-n_r; |m|+1; \mu r^2\right) \exp\left(-\mu r^2/2 +i m\vf\right), 
\quad \mu \equiv M\omega/\hbar.
\label{psinm}
\ee
Here $m=0, \pm 1, \pm 2, \ldots$, $n_r = 0,1,2,\ldots$
and $\Phi(a;c;z)$ is the confluent hypergeometric function.
Function (\ref{psinm}) can be written also in terms of the generalized Laguerre polynomials. Using the definition \cite{Bateman,Lebedev}
\[
\Phi(-n;\alpha;z) = \frac{n!\Gamma(\alpha+1)}{\Gamma(n+\alpha+1)}L_n^{(\alpha)}(z)
\]
we obtain the following expression for normalized solutions (frequently called, especially in the optics applications, as
Laguerre--Gauss functions):
\be
\fl
\psi_{n_r m}(r,\vf) = \sqrt{\frac{\mu n_r!}{\pi \left(n_r +|m|\right)!}}
\left(\mu r^2\right)^{|m|/2}L_{n_r}^{(|m|)}\left(\mu r^2\right) \exp\left(-\mu r^2/2 +i m\vf\right).
\label{psiLag}
\ee

From the well known form of energy spectrum 
\be
E_{n_r m} =\hbar\omega\left(1 +|m| +2n_r\right)
\label{Eosc}
\ee
it is clear that the minimal energy {\em eigenvalue\/} for a fixed {\em eigenvalue\/} of the angular momentum 
$\hbar m$ equals $E_{min}(m)=\hbar\omega\left(1 +|m|\right)$.
But what is the minimal {\em mean\/} value of energy for {\em superpositions\/} of energy eigenstates with a fixed {\em mean\/} 
value of angular momentum $\hbar {\cal L}$ and an {\em arbitrary\/} value of ${\cal L}$?
Taking into account the orthogonality of functions $\psi_{n_r m}$ one can see that the answer is 
\be
\langle E\rangle_{min}({\cal L}) =  \hbar\omega(1+  |{\cal L}|).
\label{Emin1}
\ee
This value is achieved, for example, in superpositions of states with zero value of radial quantum number $n_r$ and {\em the same signs\/}
of quantum numbers $m$:
\be
\fl
\psi_{min} = \sum_{m} c_m \psi_{0m}, \quad \sum_{m} |c_m|^2=1, \quad \sum_{m} m|c_m|^2 =\hbar{\cal L},
\quad \sum_{m} |m|\,|c_m|^2 =\hbar|{\cal L}|.
\label{pack1}
\ee
Obviously, the number of possible superpositions of this kind is infinite. We wish to know, if there exist {\em Gaussian packets\/}
satisfying equation (\ref{Emin1})? The  answer to this question is positive, but it is not obvious or trivial, 
as we show in the subsequent sections. In particular, the result depends crucially on the mutual directions of  the `external'
rotation (related to the motion of the center of packet) and the `internal' one (related to the evolution of quantum fluctuations
and the direction of rotation of the ellipse of constant probability density).
Note that the intensity (probability density) of the eigenfunctions  $|\psi_{0m}|^2$ with $m \neq 0$  equals zero at the center.
In contrast, Gaussian packets with nonzero mean angular momentum have the maximal probability density at the center.
Moreover, it will be shown in section \ref{sec-osc} that the minimizing packets maintain their shape and size
in the coordinate space during the time evolution (for a nonzero oscillator potential or nonzero magnetic field), 
rotating like a rigid body, but being {\em squeezed\/} at the same time.
This is a generalization of Schr\"odinger's packets \cite{Schrodpac} that gave rise eventually to the concept of coherent states.
Therefore the family of rotating minimum energy packets is a distinguished subfamily of all Gaussian states in two spatial dimensions,
which deserves the detailed analysis.

The plan of the paper is as follows. Section \ref{sec-Gauss}
is devoted to general properties of two-dimensional Gaussian packets.
The problem of energy minimization under the constraint of the fixed mean angular momentum is solved in section \ref{sec-osc}.
The statistical properties (such as squeezing, in particular) of the extremal packets and their evolution in time are considered
in that section as well.
In section \ref{sec-fluct} we calculate variances of the energy and angular momentum. The expansion coefficients over the
Laguerre--Gauss eigenstates  (\ref{psiLag}) are found in section \ref{sec-decompos}. The difference between `co-rotating' and
`anti-rotating' packets becomes especially clear in quite different expressions for these coefficients. In particular, we show
that `co-rotating' packets with specific values of the `internal' and `external' angular momenta possess more narrow distributions
in the Fock space than the Poissonian distribution.
Section \ref{sec-magosc} is devoted to the minimal energy Gaussian packets for a charged oscillator 
and a charged free particle in a homogeneous magnetic field.
In section \ref{sec-free} we study the peculiarities of the rotating Gaussian packets describing a free quantum particle.

\section{Gaussian packets in two dimensions}
\label{sec-Gauss}

We consider normalized Gaussian packets 
in two dimensions
\be
\psi(x,y) = \tilde{N} \exp\left[-\mu\left( a x^2 +b xy + c y^2\right) +Fx +Gy\right]
\label{psi}
\ee
where $\mu$ is a constant scale factor. It is convenient to choose $\mu=M\omega/\hbar$ in the case of isotropic harmonic oscillator,
but in other cases this factor can take different values.
In polar coordinates $x=r\cos(\vf)$ and $y=r\sin(\vf)$ the packet has the form 
\beqn
\psi(r,\vf)&=& \tilde{N} \exp\Big\{ -\,\frac{\mu}{2} r^2\left[ a+c + (a-c)\cos(2\vf) +b \sin(2\vf) \right]
\nonumber \\ &&
+ r\left[F\cos(\vf) + G\sin(\vf) \right]\Big\}.
\label{G-pol}
\eeqn

 It is useful to separate real and imaginary parts of five complex coefficients 
\be
\fl
a =\alpha/2 +i\chi_a, \quad b=\beta +i\rho, \quad c=\gamma/2 +i\chi_c, \quad F=F_1+iF_2, \quad G=G_1 +iG_2.
\label{abc}
\ee
Then the probability density has the form
\be
\fl
|\psi^2|(x,y) =|N|^2  \exp\left[-\mu\left( \alpha \tilde{x}^2 +2\beta \tilde{x}\tilde{y} + \gamma \tilde{y}^2\right)\right], \quad
\tilde{x} = x-x_0, \quad \tilde{y} = y-y_0, 
\label{psi2}
\ee
where
\be
\fl
|N|^2 =|\tilde{N}|^2 \exp\left[\mu\left( \alpha {x}_0^2 +2\beta {x}_0{y}_0 + \gamma {y}_0^2\right)\right]
=  \mu\sqrt{\Delta}/\pi, \quad \Delta = \alpha\gamma - \beta^2, 
\label{N}
\ee
\be
x_0=\left(\gamma F_1 - \beta G_1\right)/(\mu\Delta), \quad
y_0=\left(\alpha G_1 - \beta F_1\right)/(\mu\Delta).
\label{x0y0}
\ee
Obviously $\alpha,\gamma >0$, whereas all other real parameters can assume any sign, obeying the only restriction
$\beta^2 <\alpha\gamma$. Parameters $x_0$ and $y_0$ are coordinates of the center of packet. They coincide with 
mean values of coordinates: $x_0=\langle \hat{x}\rangle$, $y_0=\langle \hat{y}\rangle$.

The lines of constant relative probability density, defined by the equation $|\psi(x,y)|^2=\exp(-\nu)|\psi(x_0, y_0)|^2$,
 are ellipses, whose 
 major/minor semi-axes $a_{\pm}$, eccentricity $\vep$ and area $Y$ are given by the formulas
\be
\fl
a_{\pm}^2 = \frac{2\nu}{\alpha+\gamma \mp R}, \quad \varepsilon^2 = \frac{2R}{\alpha+\gamma +R}, \quad
Y= \frac{\pi\nu}{\sqrt{\Delta}} , \qquad  R = \sqrt{(\alpha-\gamma)^2 +4\beta^2}.
\label{apm}
\ee
The angle $\theta$ between the directions of the major/minor axes and the coordinate axes can be found from the equation
\be
\tan(2\theta) ={2\beta}/(\gamma-\alpha).
\label{theta}
\ee
Mean values of momenta $p_{x0}=\langle \hat{p}_x\rangle$ and $p_{y0}=\langle \hat{p}_y\rangle $ are 
\be
p_{x0} = \hbar\left[F_2 -\mu\left(2\chi_a x_0 + \rho y_0\right)\right], \quad
p_{y0} = \hbar\left[G_2 -\mu\left(2\chi_c y_0 + \rho x_0\right)\right].
\label{pxpy}
\ee
The probability current density vector has the components

\be
\left(j_x,j_y\right) = (\hbar/m)|\psi|^2\left(F_2 -\mu\left[\rho y + 2\chi_a x\right],  G_2 -\mu\left[\rho x + 2\chi_c y\right]\right).
\label{j}
\ee

Let us introduce the notation
$\overline{AB} \equiv \langle \hat{A}\hat{B}+\hat{B}\hat{A}\rangle/2 -\langle \hat{A}\rangle\langle\hat{B}\rangle$ 
for the symmetrical covariances of operators $\hat{A}$ and $\hat{B}$.  
The following expressions hold for covariances of the coordinates and momenta operators:
\be
\overline{x^2} = \gamma/(2\mu\Delta), \quad \overline{y^2} = \alpha/(2\mu\Delta), \quad
\overline{xy} = -\beta/(2\mu\Delta),
\label{ovx2}
\ee
\be
\overline{p_x^2} = \mu\hbar^2\left[\gamma\left(\alpha^2 +4\chi_a^2\right) +\alpha\left(\rho^2 - \beta^2\right)
-4\beta\rho\chi_a\right]/(2\Delta) , 
\ee
\be
\overline{p_y^2} = \mu\hbar^2\left[\alpha\left(\gamma^2 +4\chi_c^2\right) +\gamma\left(\rho^2 - \beta^2\right)
-4\beta\rho\chi_c\right]/(2\Delta) ,
\label{ovpy2}
\ee
\be
\overline{xp_x} = \hbar\left(\beta\rho -2\gamma\chi_a\right)/(2\Delta), \quad 
\overline{yp_y} = \hbar\left(\beta\rho -2\alpha\chi_c\right)/(2\Delta).
\ee
\be
\overline{p_x p_y} = \mu\hbar^2\left[ \beta\left(\Delta -\rho^2 - 4\chi_a\chi_c\right)
+2\rho\left(\alpha\chi_c +\gamma\chi_a\right)\right]/(2\Delta),
\ee
\be
\overline{xp_y} = \hbar\left(2\beta\chi_c -\rho\gamma\right)/(2\Delta), \quad 
\overline{yp_x} = \hbar\left(2\beta\chi_a -\rho\alpha\right)/(2\Delta).
\ee
The mean value of the angular momentum operator 
$\hat{L}_z = \hat{x}\hat{p}_y - \hat{y}\hat{p}_x$ can be written as
$\langle \hat{L}_z\rangle \equiv \hbar{\cal L} = \hbar\left({\cal L}_c + {\cal L}_i\right)$, where the `classical' (related to the
motion of the packet center) and `intrinsic' (related to the quantum fluctuations) parts are given by the following expressions:
\be
\hbar{\cal L}_c = x_0 p_{y0} - y_0 p_{x0} ,
\label{Lc}
\ee
\be
{\cal L}_i  = \left(\overline{xp_y} -\overline{yp_x}\right)/\hbar 
= \left[2\beta\left(\chi_c-\chi_a\right) + \rho(\alpha-\gamma)\right]/(2\Delta).
\label{L}
\ee
It is important that the values of ${\cal L}_c$ and ${\cal L}_i$ are totally independent for all Gaussian packets.
Moreover, only packets exhibiting some asymmetry in their shapes ($a\neq c$) can possess a nonzero intrinsic mean angular momentum.
Our goal is to find families of the `best' packets satisfying some additional requirements.

\section{Minimal energy packets of the harmonic isotropic oscillator}
\label{sec-osc}

Let us suppose that (\ref{psi}) is the wave function of a particle with mass $M$ moving in the  isotropic harmonic potential
$M\omega^2\left(x^2+y^2\right)/2$. Then it is convenient to choose the scale factor as $\mu = {M\omega}/{\hbar}$.  
The mean energy ${\cal E}$, as well as the mean angular momentum, is the sum of two independent terms:
${\cal E}= {\cal E}_c + {\cal E}_i$, where
\be
{\cal E}_c = \frac1{2M}\left(p_{x0}^2 + p_{y0}^2\right) 
+ \frac{M\omega^2}{2}\left(x_0^2 + y_0^2\right)
\label{Ec}
\ee
and 
\[
{\cal E}_i = \frac1{2M}\left(\overline{p_{x}^2} + \overline{p_{y}^2}\right) 
+ \frac{M\omega^2}{2}\left(\overline{x^2} + \overline{y^2}\right).
\]
Using equations (\ref{ovx2})-(\ref{ovpy2}) we can write
\be
{\cal E}_i =\frac{\hbar\omega}{4\Delta} \left[
(\alpha+\gamma)\left(1+\Delta +\rho^2\right) +4\left(\gamma\chi_a^2\ +\alpha\chi_c^2\right) 
 -4\beta\rho\left(\chi_a +\chi_c\right)\right].
\label{E}
\ee
Obviously the quantities ${\cal E}_{c}$ and ${\cal E}_{i}$ are totally independent.

The minimization of the `classical' energy ${\cal E}_{c}$ with the fixed value of `classical' angular momentum ${\cal L}_{c}$ 
(\ref{Lc}) can be done easily:
${\cal E}_{c}^{(min)}({\cal L}_c)= \hbar\omega|{\cal L}_{c}|$.
The minimizing trajectories of the packet center are circles:
\be
  x_0^2 + y_0^2 =  R^2 = |{\cal L}_{c}|/\mu, \quad
p_{x0} = - \lambda_c M\omega y_0,  \quad p_{y0} =  \lambda_c M\omega x_0, 
\label{circle}
\ee
where coefficient $\lambda_c=\pm1$ determines the direction of rotation: ${\cal L}_c =\lambda_c |{\cal L}_c|$.

To find the minimal mean  value of  `intrinsic' energy ${\cal E}_i$ for the fixed mean value of `intrinsic' angular momentum ${\cal L}_i$ 
(\ref{L}) we introduce the new set of parameters 
\be
\alpha+\gamma =2g, \quad \alpha-\gamma =2\xi, \quad \chi_a-\chi_c = 2\chi, \quad  \chi_a+\chi_c = 2z,
\label{algam-g}
\ee
\be
\alpha = g+\xi, \quad \gamma =g-\xi, \quad \chi_a=z +\chi, \quad \chi_c = z-\chi
\ee
and solve equation (\ref{L}) with respect to $\chi$:
\be
\chi =  (\rho\xi -{\cal L}_i \Delta)/(2\beta).
\label{chi-rho}
\ee
Putting (\ref{algam-g})-(\ref{chi-rho}) into (\ref{E}) we can write 
${\cal E}_i = E_1 +E_2$, where
\be
E_1(g,\eta) =\frac12\hbar\omega g \left[ 1 + \frac1{\Delta} + \frac{{\cal L}_i^2 \Delta}{\eta^2}\right], 
\label{E1}
\ee
\be
E_2 = \frac12\hbar\omega \left[\frac{4z^2}{g} +\frac{g\eta^2}{\beta^2 \Delta}
\left( \rho - \frac{\xi{\cal L}_i\Delta}{\eta^2} - \frac{2z\beta}{g}\right)^2
\right],
\ee
\be
\eta^2 = \xi^2 +\beta^2, \qquad \Delta = g^2 -\eta^2.
\ee
The minimum of $E_2$ is obviously achieved for
\be
z=0, \qquad \rho = \xi{\cal L}_i\Delta/\eta^2.
\label{z0}
\ee
Therefore the minimum of ${\cal E}_i$ coincides with the minimum of function $E_1$ (\ref{E1}),
 which is achieved at (see \ref{ap1} for details)
\be
g=1, \qquad \eta^2 =\frac{|{\cal L}_i|}{1+ |{\cal L}_i|}, 
\label{g1}
\ee
so that
\be 
{\cal E}_i^{(min)} =  \hbar\omega(1+  |{\cal L}_i|).
\label{Emin}
\ee
Consequently, the total minimal mean energy equals
\be
{\cal E}_{min} = \hbar\omega\left(1 +|{\cal L}_{i}| +|{\cal L}_{c}|\right).
\label{Etotmin}
\ee
If the signs of ${\cal L}_{i}$ and ${\cal L}_{c}$ coincide, then ${\cal E}_{min} =  \hbar\omega(1+  |{\cal L}|)$
in accordance with  (\ref{Emin1}). But the mean energy can be much bigger in the case of opposite
directions of `internal' and `external' rotations.

We see that the minimizing states are degenerate (this is not surprising for the isotropic oscillator), since the same
values of energy and angular momentum (\ref{Emin}) are achieved for the Gaussian packets with the following real
coefficients (we assume $\eta \ge 0$ in all formulas below):
\be
\alpha=1+\eta\cos(u), \quad \gamma = 1-\eta\cos(u), \quad \chi=-\lambda\eta\sin(u)/2,
\ee
\be
\beta = \eta\sin(u), \quad \rho = \lambda \eta\cos(u), 
\label{coef}
\ee
Here $\lambda=\pm 1$ is responsible for the sign of mean intrinsic angular momentum 
(${\cal L}_i>0$ for $\lambda=1$ and ${\cal L}_i<0$ for $\lambda=-1$)
and $u$ is an arbitrary phase. An additional obvious degeneracy is related to the choice of parameters $x_0$ and $y_0$.
Complex coefficients of function (\ref{psi}) have the form (with $x_0=R\cos(v)$ and $y_0=R\sin(v)$)
\be
\fl
a=\frac12\left[1 +\eta\exp(-i\lambda u)\right], \quad
c=\frac12\left[1 -\eta\exp(-i\lambda u)\right], \quad
b= i\lambda\eta\exp(-i\lambda u),
\label{compl}
\ee
\be
F= \mu R \left\{\exp\left(-i\lambda_c v\right) + \eta\exp\left[i\lambda (v-u)\right]\right\},
\ee
\be
G= i\mu R \left\{\lambda_c\exp\left(-i\lambda_c v\right) + \lambda\eta\exp\left[i\lambda (v-u)\right]\right\}.
\label{G}
\ee

Combining equation  (\ref{G-pol}) with (\ref{x0y0}), (\ref{pxpy}), (\ref{circle}) and (\ref{compl}) we obtain the specific form 
of the minimal energy Gaussian packets in the polar coordinates:
\beqn
\psi_{min}(r,\vf) &=& \sqrt{\mu/\pi}(1-\eta^2)^{1/4} \exp\left\{ -\,\frac{\mu}{2} r^2\left[ 1 + \eta \exp(2i\lambda\vf -i\lambda u) \right]
\right. \nonumber \\   && \left. +
\mu Rr \left(\exp\left[i\lambda_c(\vf - v)\right] +\eta\exp[i\lambda(\vf +v -u)]\right) -\frac{\Phi}{2}
\right\}
\label{G-polmin}
\eeqn
where
\be
\Phi = \left|{\cal L}_c\right|\left[1 +\eta\cos(u-2v)\right].
\ee

\subsection{The second-order statistical moments and squeezing coefficients}

The covariances of coordinates and momenta in the minimum energy Gaussian packets have the following values:
\be
\overline{x^2} = \frac{\hbar(1+  |{\cal L}_i|)}{2M\omega}\left(1-\eta\cos(u)\right), 
\quad \overline{y^2} = \frac{\hbar(1+  |{\cal L}_i|)}{2M\omega}\left(1+\eta\cos(u)\right),
\label{x2}
\ee
\be
\fl
\overline{p_x^2} = \frac12{M\omega\hbar}(1+  |{\cal L}_i|)\left(1+\eta\cos(u)\right),  \quad  
\overline{p_y^2} = \frac12{M\omega\hbar}(1+  |{\cal L}_i|)\left(1-\eta\cos(u)\right), 
\ee
\be
\overline{xp_x} = \frac{\hbar}{2}(1+  |{\cal L}_i|)\lambda \eta\sin(u), \quad 
\overline{yp_y} = -\,\frac{\hbar}{2}(1+  |{\cal L}_i|)\lambda \eta\sin(u),
\ee
\be
\overline{xy} = -\, \frac{\hbar(1+  |{\cal L}_i|)}{2M\omega}\eta\sin(u), \quad
\overline{p_x p_y} = \frac12{M\omega\hbar}(1+  |{\cal L}_i|)\eta\sin(u),
\ee
\be
\overline{xp_y} =\frac12 \hbar{\cal L}_i\left[1-\cos(u)/\eta\right], \quad
\overline{yp_x} = -\,\frac12 \hbar{\cal L}_i\left[1+\cos(u)/\eta\right].
\label{xpy}
\ee

We see that the partial intrinsic energies ${\cal E}_i^{x}$ and ${\cal E}_i^{y}$ coincide, as well as the Robertson--Schr\"odinger
uncertainty products $U_x$ and $U_y$, where $U_x \equiv \overline{x^2}\,\overline{p_x^2} -\left(\overline{xp_x}\right)^2$:
\be
U_x=U_y =  \frac{\hbar^2}{4}(1+  |{\cal L}_i|).
\label{UxL}
\ee
Let us define the correlation coefficient between variables $g$ and $f$ as 
$r_{gf}=\overline{gf}/\sqrt{\overline{g^2}\,\overline{f^2}}$.
Then
\be
r_{p_xp_y}=-r_{xy}=\lambda r_{xp_x} = -\lambda r_{yp_y} = \frac{\eta\,\sin(u)}{\sqrt{1-\eta^2\cos^2(u)}}.
\ee
The case of $u=0$ (pure imaginary coefficient $b$ and maximally different real coefficients $a$ and $c$) corresponds
to the absence of correlations in four pairs of variables shown above. Another extreme case of $u=\pi/2$
(real coefficient $b$ and complex conjugate coefficients $a=c^*$) corresponds to the maximal correlation coefficients
in the same pairs:
$|r_{max}|=\eta = \sqrt{ |{\cal L}|/(1+ |{\cal L}|)}$.

The best characteristics of {\em squeezing\/} in the $x$-mode is the {\em invariant squeezing coefficient\/} \cite{GSS,R1}
(or `principal squeezing' \cite{Luks})
\be
S_x = 2\left(\tilde{\cal E}_x - \sqrt{\tilde{\cal E}_x^2 -\tilde{U}_x}\right), \qquad \tilde{\cal E}={\cal E}/(\hbar\omega),
\quad \tilde{U}=U/\hbar^2.
\ee
For the states under study we have
\be
S_x=S_y = \frac{1}{1+\eta} < 1,
\ee
so that the states are squeezed, although the concrete `directions' of squeezing can be different. For example, for $u=0$ 
we see squeezing in the $x$-coordinate and $p_y$-momentum.
However, the maximal degree of squeezing cannot exceed 50\%, since $\eta^2<1$.

\subsection{Time evolution of packets}

The origin of the degeneracy of minimizing states with respect to arbitrary phases $u$ and $v$ becomes clear if one considers the
time evolution of these states. It is given by the integral
\[
\psi(x,y;t) =\int G(x,y;x',y';t)\psi(x',y';0) dx' dy',
\]
where the propagator $G(x,y;x',y';t)$ for the two-dimensional isotropic harmonic oscillator (calculated for the first time by Kennard \cite{Kenn}) reads
(here ${\bf r} =(x,y)$)
\be
\fl
G({\bf r},{\bf r}';t) =\frac{\mu}{2\pi i \sin(\omega t)} \exp\left\{\frac{i\mu}{2\sin(\omega t)}\left[
\cos(\omega t)\left({\bf r}^2 +{\bf r}'^2 \right) -2{\bf r}{\bf r}'\right]\right\}.
\label{Green}
\ee
Performing the integration with initial function (\ref{psi}), parametrized as in (\ref{compl})-(\ref{G}) with initial phases $u_0$
and $v_0$,
one can find that function $\psi(x,y;t)$ has the same form, with the only difference that $u_0$ and $v_0$ should be
replaced by time dependent phases 
\be
u(t)=u_0 +2\lambda\omega t, \quad v(t)=v_0 + \lambda_c \omega t.
\label{uvt}
\ee

This means that the lines of constant probability density are ellipses rotating around the central points $(x_0(t), y_0(t)) $
with the angular velocity $2\lambda\omega$ without changing
their shapes (like the ellipses of constant quasiprobability in the phase plane $xp$ of one-dimensional harmonic oscillator). The minor
axis is inclined by angle $u(t)/2=u_0/2 +\lambda\omega t$ with respect to $x$-axis. The major
and minor axes of the ellipse are proportional to $\left(1 \mp \eta\right)^{-1/2}$ (with equal scaling factors), and the 
ellipse eccentricity equals $\vep= \left[2\eta/\left(1+\eta\right)\right]^{1/2}$.

\subsection{Verification of universal invariants}

If one combines all coordinate and momentum (co)variances in the symmetric matrix $Q$, then the quantities ${\cal D}_m$
defined according to the expansion
\be
\mbox{det}\left(Q - \gamma \Sigma\right) = \sum_{m=0}^{2n} {\cal D}_m \gamma^m
\label{univ}
\ee
do not depend on time for any quadratic Hamiltonian. Here $\Sigma$ is the antisymmetric matrix constructed from the $c$-number
commutators between coordinates and momenta operators, $\gamma$ is an arbitrary auxiliary parameter and $2n$ the dimensionality
of matrices $Q$ and $\Sigma$. Quantities ${\cal D}_m$ were named {\em universal quantum invariants\/} in
\cite{183vol,univ85,univ00,univOl}. They are also called as {\em symplectic invariants\/} \cite{Simon94}
(or {\em characteristic invariants\/} \cite{Sud95}), especially in the
quantum information literature \cite{Seraf04,Ades05,Seraf07,Weed12}.
In the one-dimensional case ($2\times2$ matrices $Q$ and $\Sigma$) the only nontrivial invariant $\mbox{det}(Q)$ coincides with
the Robertson--Schr\"odinger uncertainty product $U_x$, and its value for minimal energy packets is given by equation (\ref{UxL}).
In the two-dimensional case ($4\times4$ matrices $Q$ and $\Sigma$) there exist two invariants. 
One of them is again  ${\cal D}_0 \equiv \mbox{det}(Q)$.
One can verify that the set of (co)variances (\ref{x2})-(\ref{xpy}) results in the time-independent value of ${\cal D}_0$. Moreover,
this value does not depend on the angular momentum: ${\cal D}_0=\hbar^4/16$. Actually, this is the common value for  all
{\em pure Gaussian states}, since such states minimize the generalized uncertainty relation ${\cal D}_0\ge \hbar^4/16$
\cite{183vol,univ85,Sud95,Rob34}.
The second invariant of the two-dimensional systems  
\be
{\cal D}_2/\hbar^2= 
\left(\overline{yp_y}\right)^2 +\left(\overline{xp_x}\right)^2 +2\,\overline{xp_y}\;\overline{yp_x}
-2\overline{xy}\;\overline{p_xp_y} -\overline{p_x^2}\;\overline{x^2} - \overline{p_y^2}\;\overline{y^2}
\ee
also does not depend on the angular momentum for the set (\ref{x2})-(\ref{xpy}): ${\cal D}_2= -\hbar^4/2$. Therefore
the combination ${\cal D}_0 + {\cal D}_2/4 + \hbar^4/16$ equals zero, which is the minimal possible value according
to another generalized uncertainty relation \cite{183vol,univ85}.

In the current quantum information studies the important quantities are so called `symplectic eigenvalues' of the covariance matrix,
defined as eigenvalues of matrix $\Sigma^{-1}Q$. It is known that these  eigenvalues consist of $n$ pairs 
$(\kappa_1, -\kappa_1) \ldots (\kappa_n, -\kappa_n)$. Their connection with the universal invariants is seen from the identity
\be
\mbox{det}\left(Q - \gamma \Sigma\right) =\mbox{det} (\Sigma) \mbox{det}\left(\Sigma^{-1}Q - \gamma I_{2n}\right)
= \mbox{det} (\Sigma) \prod_{j=1}^{n}\left(\gamma^2 -\kappa_j^2\right),
\ee
where $I_{2n}$ is the $2n\times2n$ identity matrix. We have $\mbox{det} (\Sigma)=\hbar^4$ in the two dimensional case involved
($n=2$). 
Consequently ${\cal D}_0 =\hbar^4 \kappa_1^2 \kappa_2^2$ and ${\cal D}_2 = -\hbar^4 \left(\kappa_1^2 +\kappa_2^2\right)$.
This means that the minimum energy states with covariances given by equations (\ref{x2})-(\ref{xpy}) possess the minimal possible 
symplectic eigenvalues $|\kappa_1|=|\kappa_2| =1/2$, which do not depend on the mean angular momentum value ${\cal L}_i$.

\section{Energy and angular momentum fluctuations}
\label{sec-fluct}

It is interesting to know the energy and angular momentum variances $\sigma_E =\langle \hat{H}^2\rangle - \langle \hat{H}\rangle^2$
and $\sigma_L =\langle \hat{L}^2\rangle - \langle \hat{L}\rangle^2$. One approach is to calculate the fourth order moments
of coordinates and momenta. This can be done relatively easy for the Gaussian states, because their Wigner functions 
$W\left(x,y,p_x,p_y\right)$ are also Gaussian,
so that one can use classical formulas (with some modifications due to the non-commutativity of the coordinate and momentum operators) 
for average values of the Gauss distributions (see, e.g. \cite{183vol}). 
Using the representation of quadrature operators as sums of average and fluctuating parts, e.g., $\hat{x}=x_0 + \tilde{x}$,
$\hat{p}_x =p_{x0} +\tilde{p}_x$ and so on, we can write
\beqn
\langle \hat{L}^2\rangle &=& \left(x_0 p_{y0} -y_0 p_{x0}\right)^2 
+2x_0 y_0\left[ M\omega \lambda_c \left(\overline{yp_y} -\overline{xp_x}\right) +(M\omega)^2 \overline{xy}
-\overline{p_x p_y} \right]
\nonumber \\ &&
+ 2  M\omega \lambda_c \left[ \overline{x p_y}\left(2x_0^2 +y_0^2\right)
- \overline{y p_x}\left(2y_0^2 +x_0^2\right)\right]
\nonumber \\&&
+ \langle  \tilde{ x}^2\tilde{ p}_y^2 + \tilde{ y}^2\tilde{ p}_x^2 - \tilde{ x}\tilde{ p}_y\tilde{ y}\tilde{ p}_x
- \tilde{ y}\tilde{ p}_x\tilde{ x}\tilde{ p}_y \rangle.
\label{L2full}
\eeqn
We take into account that mean values of products of any {\em three\/} operators marked with tildes are equal to zero for
Gaussian states.
The average values appearing in the first and second lines of (\ref{L2full}) are given by formulas (\ref{x2})-(\ref{xpy}).
To calculate the fourth-order {\em central\/} moments contained in the last line, we use the known formula for Gaussian states,
connecting mean values of {\em symmetrical\/} (or Wigner--Weyl) products \cite{Hill84} of four operators $\hat{A}$, $\hat{B}$, $\hat{C}$
and $\hat{D}$ (with zero mean values) and sums of pair products of their covariances \cite{183vol}
\beqn
\langle{ABCD}\rangle_W & \equiv& \int W\left(x,y,p_x,p_y\right)\,ABCD\,dxdydp_xdp_y/(2\pi\hbar)^2
\nonumber \\ & =&
\overline{AB}\cdot\overline{CD} + \overline{AC}\cdot\overline{BD} + \overline{AD}\cdot\overline{BC}.
\label{basic}
\eeqn
Here $A,B,C,D$ can be any of variables $x,y,p_x,p_y$. The meaning of symbol $\langle{ABCD}\rangle_W$ is the following:
this is the quantum mechanical mean value of the sum of $4!=24$ products of operators $\hat{A},\hat{B},\hat{C},\hat{D}$
taken in all possible orders, divided by the number of terms. Mean values of concrete products of operators in predefined
orders can be expressed in terms of symmetrical mean values with the aid of commutation relations.
In our case the following relations are useful (they are valid for {\em Gaussian\/} states):
\[
\langle \hat{x}^2\hat{p}_y^2\rangle = 2\left(\overline{xp_y}\right)^2 + \overline{x^2}\cdot\overline{p_y^2},
\]
\[
\fl
\langle \hat{x}\hat{p}_y\hat{y}\hat{p}_x + \hat{y}\hat{p}_x\hat{x}\hat{p}_y\rangle 
= 2\langle{xyp_xp_y}\rangle_W +\hbar^2/2 
= 2\left(\overline{xy}\cdot\overline{p_xp_y} + \overline{xp_x}\cdot\overline{yp_y} + \overline{xp_y}\cdot\overline{yp_x}\right) +\hbar^2/2.
\]
After some algebra one can arrive at the following expression for the angular momentum variance in terms of the `external'
and `intrinsic' mean values ${\cal L}_c$ and ${\cal L}_i$:
\be
\fl
\sigma_L/\hbar^2 =  \left|{\cal L}_c\right| + 2\left|{\cal L}_i\right| \left(1+\left|{\cal L}_i\right|\right)
+\left(1+\lambda\lambda_c\right) \left|{\cal L}_c\right| \left[ \left|{\cal L}_i\right| 
 -  \sqrt{\left|{\cal L}_i\right| \left(1+\left|{\cal L}_i\right|\right)}\cos(2w) \right],
\label{sigL+-}
\ee
where
\be
w=\lambda(v-u/2) = w_0 + \left(\lambda\lambda_c -1\right)\omega t.
\label{w}
\ee
We see that the result depends on the product $\lambda\lambda_c =\pm 1$, which is positive in the case of `co-rotation'
of the packet center and ellipse axes and negative for `anti-rotating' packets.
The phase difference $w$ does not influence the angular momentum variance (as well as its mean value) in the 
`anti-rotating' case:
\be
\sigma_L/\hbar^2 = \left|{\cal L}_c\right| + 2\left|{\cal L}_i\right| \left(1+\left|{\cal L}_i\right|\right),  \quad
\lambda\lambda_c=-1.
\label{sigLa}
\ee
But this phase is important in the case of `co-rotation' (let us assume that ${\cal L}_i>0$):
\be
\fl
\sigma_L/\hbar^2 = {\cal L} +{\cal L}_i (1+ 2{\cal L}) - 2{\cal L}_c \sqrt{{\cal L}_i \left(1+{\cal L}_i\right)}\cos(2w),
  \quad
\lambda\lambda_c=+1, \quad {\cal L} ={\cal L}_i + {\cal L}_c.
\label{sigLc}
\ee
The peculiarity of this case is analyzed in section \ref{sec-compet}.

The variance of energy can be calculated in the same manner. Here we need the formulas
\[
\overline{x^4} =3\left(\overline{x^2}\right)^2, \quad 
\overline{x^2y^2}= 2\left(\overline{xy}\right)^2 + \overline{x^2}\cdot\overline{y^2},
\]
\[
\langle \hat{x}^2\hat{p}_x^2 + \hat{p}_x^2 \hat{x}^2 \rangle = 2\overline{x^2p_x^2} -\hbar^2
=4\left(\overline{xp_x}\right)^2 + 2  \overline{x^2}\cdot\overline{p_x^2} -\hbar^2.
\]
The final result coincides exactly with (\ref{sigL+-}):
\be
\fl
\sigma_E/(\hbar\omega)^2 =  \left|{\cal L}_c\right| + 2\left|{\cal L}_i\right| \left(1+\left|{\cal L}_i\right|\right)
+\left(1+\lambda\lambda_c\right) \left|{\cal L}_c\right| \left[ \left|{\cal L}_i\right| 
 -  \sqrt{\left|{\cal L}_i\right| \left(1+\left|{\cal L}_i\right|\right)}\cos(2w) \right]
\label{sigE}
\ee

\section{Rotating Gaussian packets as superpositions of the Laguerre--Gauss energy eigenstates}
\label{sec-decompos}

It is interesting to find the coefficients of expansion
\be
\psi_{min} = \sum_{n_r,m} c_{n_r ,m}\psi_{n_r m}
\label{dec}
\ee
over the Laguerre--Gauss basis (\ref{psiLag}).
The simplest formulas correspond to the case ${\cal L}_i=\eta^2=0$, i.e., the absence of the intrinsic rotation. The state $\psi_{min}$
in this case is nothing but the two-dimensional coherent state with coherent parameters in $x$ and $y$ directions $\alpha_y =\pm i\alpha_x$.
Expanding (\ref{G-polmin}) in the Taylor series with respect to the radial variable $r$ 
\be
\psi_{min}^{(coh)}(r,\vf)= \sqrt{\frac{\mu}{\pi}} \exp\left( -\,\frac{\mu }{2}r^2 -\,\frac{\left|{\cal L}_c\right|}{2}\right) \sum_{k=0}^{\infty} 
\frac{(\mu Rr)^k}{k!}
 \exp\left[(ik\lambda_c(\vf - v) \right]
\label{G-exp0}
\ee
we see that $n_r \equiv 0$ and the distribution
over the energy eigenstates is Poissonian, as one may expect for the coherent states. The coefficients are nonzero for
the angular momentum eigenstates with the same sign of $m$ only:
\be
c_{0, k\lambda_c}= \frac{\left| {\cal L}_c \right|^{k/2}}{\sqrt{ k!}} \exp\left(-\left|{\cal L}_c\right|/2-ik\lambda_c v\right),
\quad k \ge 0.
\label{c0k}
\ee

Another simple case corresponds to packets with the fixed center at origin (${\cal L}_c=0$).
Then only {\em even\/} azimuthal quantum numbers $m=2k\lambda$ enter the expansion
\be
\fl
\psi_{min}(r,\vf)= \sqrt{\mu/\pi}(1-\eta^2)^{1/4} \exp\left( -\,\frac{\mu }{2}r^2 \right) \sum_{k=0}^{\infty} \frac1{k!}
\left( -\,\frac{\mu}{2} r^2 \eta\right)^k \exp\left[(ik\lambda(2\vf - u) \right].
\label{G-exp1}
\ee
Such a structure can be explained by the `two-photon' nature of squeezed vacuum states. Comparing (\ref{G-exp1})
with (\ref{psiLag}) we see that nonzero coefficients  in expansion (\ref{dec})  are
\be
c_{0, 2k\lambda}= (-1)^k (1-\eta^2)^{1/4}\frac{\eta^k\sqrt{ (2k)!}}{2^k k!} e^{-ik\lambda u}, \quad
\left|\frac{c_{0,2k+2}}{c_{0,2k}}\right|^2 =\eta^2\,\frac{2k+1}{2k+2}.
\label{c02k}
\ee
One can notice that coefficients (\ref{c02k}) coincide with that of the expansion of the vacuum squeezed state
of the one-dimensional harmonic oscillator over the Fock basis:
\[
|\zeta\rangle =\exp\left(\frac12\left[\zeta \hat{a}^{\dagger 2} -\zeta^* \hat{a}^2\right]\right)|0\rangle
= \left(1-|\zeta|^2\right)^{1/4}\sum_{m=0}^{\infty}\frac{\sqrt{(2m)!}}{2^m m!}\zeta^m |2m\rangle.
\]
In this case of pure `intrinsic' rotation the vacuum state gives the maximal contribution, although 
the distribution becomes rather flat for states with high mean angular momentum (if $1-\eta^2 \ll 1$ and $k\gg 1$). 

\subsection{Co-rotating packets}

In the general case of ${\cal L}_i \neq 0$ and ${\cal L}_c \neq 0$ with equal signs of these mean values ($\lambda=\lambda_c$)
we can express function (\ref{G-polmin}) as
\be
\psi_{min}(r,\vf)= \sqrt{\frac{\mu}{\pi}}(1-\eta^2)^{1/4} \exp\left( -\,\frac{\mu }{2}r^2 -\frac{\Phi}{2}\right) \sum_{k=0}^{\infty} 
\frac{(rA)^k}{k!} H_k\left(B \right) e^{ik\lambda\vf }
\label{G-gen}
\ee
where $H_k(z)$ is the Hermite polynomial \cite{Bateman} and
\be
\fl
A=  e^{-i\lambda u/2} \sqrt{{\mu\eta}/{2}}, \quad
B= \left[  \eta e^{iw} + e^{ - iw} \right] \sqrt{{\left| {\cal L}_c\right|}/(2\eta)} , 
\quad w=\lambda(v-u/2).
\ee
Consequently nonzero coefficients of expansion (\ref{dec}) have the form
\beqn
c_{0, k\lambda} &=& \frac{(1-\eta^2)^{1/4} }{\sqrt{ k!}}
\left(e^{-i\lambda u}\eta/2\right)^{k/2} 
H_k\left(\left[  \eta e^{iw} + e^{ - iw} \right] \sqrt{{\left| {\cal L}_c\right|}/(2\eta)} \right) \nonumber \\ &&
\times \exp\left\{-\left|{\cal L}_c\right|\left[1 +\eta\cos(2w)\right]/2\right\}, \quad k \ge 0.
\label{c0cor}
\eeqn
Note that phase $w$ does not depend on time for coinciding directions of the `internal' and `external' rotations.

To understand better the influence of phase $w$ on the behavior of the probabilities $p_k=\left|c_{0, k\lambda}\right|^2$, 
let us consider first the special
case of $w=0$ and ${\cal L} >0$: 
\be
p_k= \frac{(1-\eta^2)^{1/2} \eta^k}{2^k{ k!}}
H_k^2\left( (\eta +1) \sqrt{ {\cal L}_c/(2\eta)}\right)
 \exp\left[-{\cal L}_c(1 +\eta)\right] .
\label{Lic0}
\ee
For big `internal' angular momenta ${\cal L}_i \gg 1$ we consider $\eta=1-\vep$ with $\vep \ll 1$,  so that ${\cal L}_i \approx (2\vep)^{-1}$. 
Then equation (\ref{Lic0}) can be simplified if ${\cal L}_i \gg{\cal L}_c$.
Using the approximate formula $(1-\vep)^k\approx \exp(-k\vep)$  we get the expression

\be
p_k \approx  \frac{H_k^2\left(\sqrt{2{\cal L}_c } \right) }{2^k{ k!}\sqrt{{\cal L}_i}}
 \exp\left[-2{\cal L}_c  -k/(2{\cal L}_i) \right]
\ee
where contributions of `internal' and `external' rotations are factorized.
In the most interesting region $k\sim {\cal L} \gg 1$, the argument of the Hermite polynomial is much smaller than its index.
Then the known asymptotics of the Hermite polynomials 
\cite{Bateman,Lebedev} together with the Stirling formula for factorials lead to the most simple expression if
${\cal L}_i \gg{\cal L}_c \gg 1$:
\be
p_k \approx \frac{2\exp(-k/{\cal L})}{\sqrt{\pi k{\cal L}}}\cos^2\left(\sqrt{2{\cal L}_c(2k+1)}-k\pi/2\right).
\label{pk-cos}
\ee
We see that probabilities $p_k$ rapidly oscillate around some slowly decaying average distribution. Replacing $\cos^2(...)$
by its average value $1/2$ and integrating over $k$ from $0$ to $\infty$ (i.e., using the simplest form of the
Euler--Poisson summation formula) we arrive at the correct normalization $\sum_{k=0}^{\infty}p_k=1$. This shows the
reliability of the approximate formula (\ref{pk-cos}).

Now let us suppose that $w=\pi/2$, ${\cal L}_i \gg 1$ and ${\cal L}_c \ll{\cal L}_i$. Then the argument of the Hermite
polynomial in equation (\ref{c0cor}) equals approximately $-i\sqrt{{\cal L}_c/\left(8{\cal L}_i^2\right)}$, i.e, it is close to zero. The argument of
the exponential function in (\ref{c0cor}) equals approximately $-{\cal L}_c/\left(2{\cal L}_i\right)$, so that it is close to zero as well.
Consequently, the distribution $p_k$ is close to (\ref{c02k}) in this case.

It can be interesting to calculate the mean value of the angular momentum using coefficients (\ref{c0cor}). This can be done
with the aid of generating function (dependent on an auxiliary variable $z$)
\be
\fl
G(z) = \sum_{k=0}^{\infty} \left| c_{0, k\lambda}\right|^2 z^k =
\sqrt{\frac{1-\eta^2}{1-z^2 \eta^2}} \exp\left[\frac{\left|{\cal L}_c\right|(z-1)}{1-z^2 \eta^2}
\left[1-z\eta^2 +\eta(1-z)\cos(2w)\right]\right]
\ee
We used here the known Mehler formula
\[
\sum_{k=0}^{\infty}\frac{(\zeta/2)^k}{k!}H_k(x)H_k(y) = \left(1-\zeta^2\right)^{-1/2}
\exp\left[ \frac{2xy\zeta - \left(x^2 +y^2\right)\zeta^2}{1-\zeta^2}\right].
\]
The value $G(1)=1$ confirms the correct normalization of coefficients (\ref{c0cor}). The mean value of the angular momentum
can be calculated as $\hbar\sum_{k=0}^{\infty} k \left| c_{0, k\lambda}\right|^2= \hbar G'(z)\vert_{z=1}$. The result 
(for ${\cal L}_i \ge 0$ and ${\cal L}_c \ge 0$) is
$
\hbar\left[{\cal L}_c + \eta^2/(1-\eta^2)\right]= \hbar\left({\cal L}_c  + {\cal L}_i \right) = \hbar{\cal L}
$
independently of the phase $w$. 
The {\em variance\/} of the angular momentum 
$\sigma_L \equiv \langle \hat{L}_z^2\rangle - \langle \hat{L}_z\rangle^2 $
is determined by the first and second derivatives of $G(z)$ at $z=1$:
$\sigma_L/\hbar^2 = G''(1) +G'(1) - \left[G'(1)\right]^2$. 
The result coincides with equation (\ref{sigLc}).

\subsection{Competition of `external' and `intrinsic' rotations in the co-rotating case}
\label{sec-compet}

\begin{figure}[htb]
\vspace{-0.5cm}
\begin{center}
\includegraphics[height=1.0truein,angle=0]{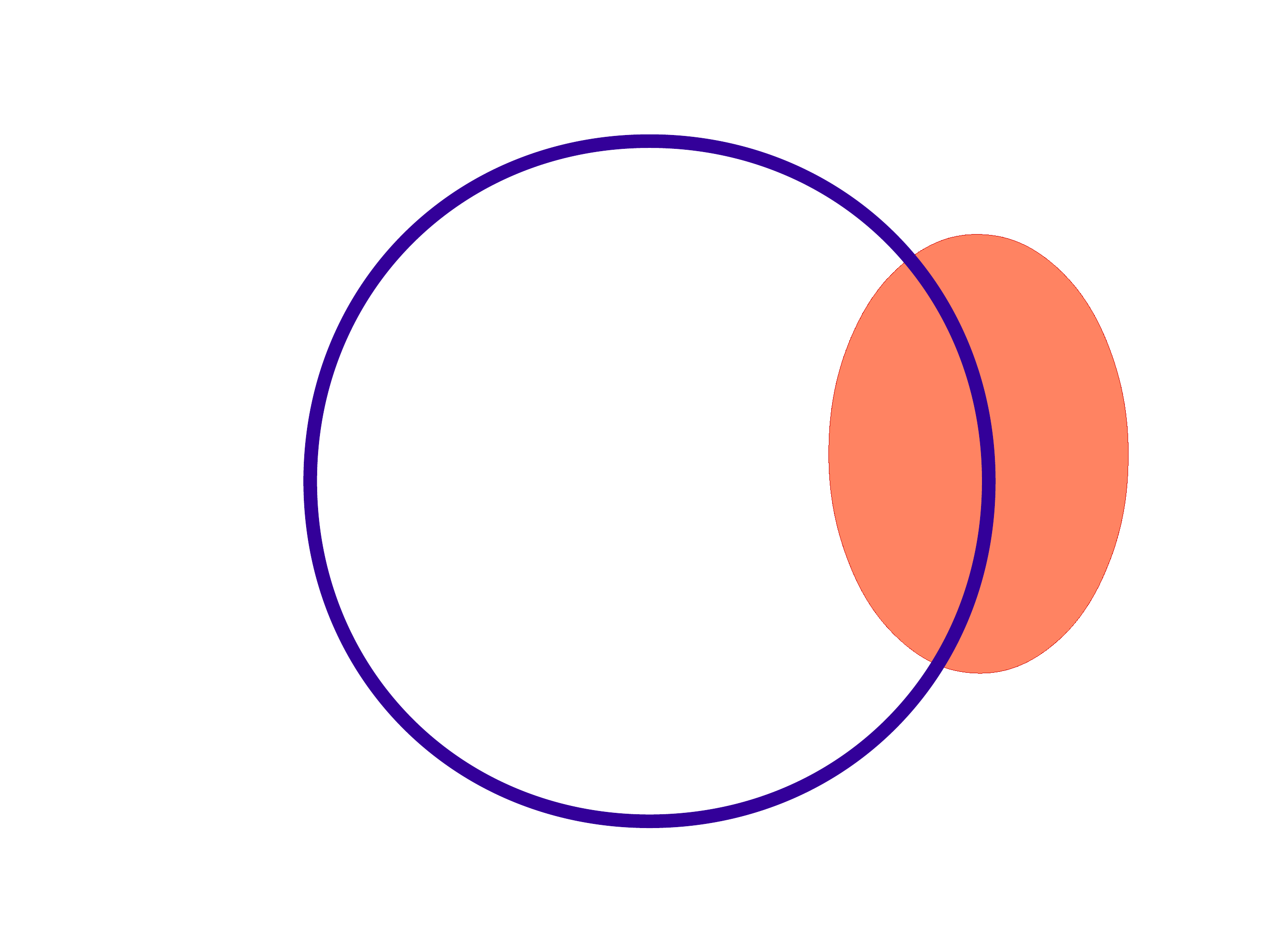}
\end{center}
\vspace{-0.5cm}
\caption{The orientation of the ellipse of constant probability density with respect to the circle which the center of
packet moves along in the case of relative angle $w=0$.  This relative configuration is maintained in time in the
co-rotating case ${\cal L}_i{\cal L}_c >0$.
}
\label{fig-circel}
\end{figure}

If ${\cal L}_i=0$ (the rotation of coherent state circular-shape packet along a circle) then we have the typical
result for the poissonian distribution $\sigma_L =\hbar^2 {\cal L}$. In the case of ${\cal L}_c=0$ we have the
typical formula for fluctuations in the squeezed vacuum state
$\sigma_L = 2\hbar^2 |{\cal L}|(1+ |{\cal L}|)$. But it is remarkable (and perhaps unexpected) that fluctuations
of the angular momentum can be smaller than that in the Poissonian distribution if both quantities ${\cal L}_i$ and
${\cal L}_c$ are different from zero (some analog of sub-Poissonian statistics). Indeed, for ${\cal L}_i \to 0$
we have $\sigma_L/\hbar^2 \approx {\cal L}\left[1- 2 \sqrt{{\cal L}_i }\cos(2w)\right] < {\cal L}$ if 
$\cos(2w)>0$. The maximal `squeezing' of the distribution $ \left| c_{0, k\lambda}\right|^2$ is observed for $w=0$,
 when the minor axis of the coordinate probability density ellipse is directed along the radius connecting the
center of packet with the origin (i.e., the ellipse is `squeezed' in the radial direction): see figure \ref{fig-circel}.

The minimum of  $\sigma_L$ with the fixed value of mean angular
momentum $\hbar{\cal L}$ and $w=0$ is achieved for ${\cal L}_i $ satisfying the equation
\be
{\cal L} = \sqrt{{\cal L}_i \left(1+{\cal L}_i\right)}\left(1+  8{\cal L}_i +8{\cal L}_i^2\right) 
+  5{\cal L}_i +12{\cal L}_i^2 + 8{\cal L}_i^3.
\label{LLi}
\ee
Combining (\ref{sigLc}) and (\ref{LLi}) we obtain the minimal variance of the total angular momentum as function
of the `intrinsic' momentum ${\cal L}_i$:
\be
\sigma_L^{min}/\hbar^2= 4{\cal L}_i \left(1+{\cal L}_i\right) 
+ \left(1+2{\cal L}_i\right) \sqrt{{\cal L}_i \left(1+{\cal L}_i\right)}.
\label{sigmin}
\ee
We see that for ${\cal L}_i \ll 1$ the minimum is achieved if ${\cal L}_i  \approx {\cal L}^2 \ll 1$. Then 
$\sigma_L^{min}/\hbar^2 \approx {\cal L} - {\cal L}^2$. The optimal ellipse eccentricity in this case equals 
$\vep_* \approx \sqrt{2{\cal L}} \ll 1$. Actually, this happens for very small values of ${\cal L}_i$. 
Increasing ${\cal L}_i$ and respectively ${\cal L}$, one can make the effect of diminishing the variance (compared
with the case of rotated coherent packet) more significant.
For example, taking 
${\cal L}_i=1/8$ we obtain ${\cal L}=13/8$,  $\sigma_L^{min}/\hbar^2 =33/32$ and $\vep_*=1/\sqrt{2}$.
For ${\cal L}_i=1/3$ we have ${\cal L}=19/3$,  $\sigma_L^{min}/\hbar^2 =26/9$ and $\vep_*=\sqrt{2/3}$.

 An approximate solution to equation (\ref{LLi}) for ${\cal L}_i \gg 1$ is 
${\cal L} \approx 16{\cal L}_i^3  $. Then $\sigma_L^{min}/\hbar^2 \approx 6{\cal L}_i^2 \approx (3/2)({\cal L}/2)^{2/3}$.
Consequently the relative minimal width of the angular momentum distribution 
\be
\sqrt{\sigma_L^{min}}/(\hbar{\cal L}) \approx \sqrt{3/2^{1/3}}(2{\cal L})^{-2/3}
\ee
can be significantly smaller than the Poissonian width ${\cal L}^{-1/2}$ in this asymptotical case, 
in spite of the small relative weight of the `intrinsic' angular momentum ${\cal L}_i/{\cal L} \approx (4{\cal L})^{-2/3}$.

\subsection{`Anti-rotating' packets}

To find the coefficients of expansion (\ref{dec}) in the case of $\lambda= -\lambda_c$ we represent the exponential function 
in the right-hand side of (\ref{G-polmin}) as the product of two sums:
\be
\psi_{min}(r,\vf)  =  \sqrt{{\mu}/{\pi}}(1-\eta^2)^{1/4} \exp\left( -{\mu }r^2/2 -{\Phi}/{2}\right)  \Psi(r,\vf),
\label{psiPsi}
\ee
\be
\Psi(r,\vf) = \sum_{k=0}^{\infty} 
\frac{(rA)^k}{k!} H_k\left(B_1 \right) e^{ik\lambda\vf } \sum_{j=0}^{\infty} \frac{(\mu Rr)^j}{j!}
e^{ij\lambda(v -\vf) },  
\label{Psi}
\ee
where
\be
 B_1 = \left(\left| {\cal L}_c\right|{\eta}/2\right)^{1/2} e^{iw}.
\label{B1}
\ee
Combining the terms with the same powers of $e^{i\lambda \vf}$ we can write
\beqn
\Psi(r,\vf) &=& \sum_{m=0}^{\infty}e^{im\lambda(\vf-u/2)}\left(\mu r^2\eta/2\right)^{m/2}
\sum_{j=0}^{\infty} \frac{\left(\mu r^2 B_1\right)^j}{j!(m+j)!}H_{m+j}\left(B_1\right) 
\nonumber \\ && +
\sum_{m=1}^{\infty}e^{-im\lambda(\vf-v)} \left(\mu r^2\left|{\cal L}_c\right|\right)^{m/2}
 \sum_{k=0}^{\infty} \frac{\left(\mu r^2 B_1\right)^k}{k!(m+k)!}H_{k}\left(B_1\right) .
\eeqn
Using the formula \cite{Lebedev} (we consider integral positive values of $m$ here)
\be
x^k = \sum_{n=0}^{k} \frac{(-1)^n k! (k+m)!}{(n+m)!(k-n)!}L_n^{(m)}(x)
\ee
we continue as
\beqn
\fl
\Psi(r,\vf) &=& \sum_{m=0}^{\infty}e^{im\lambda(\vf-u/2)}\left(\mu r^2\eta/2\right)^{m/2}
\sum_{n=0}^{\infty}\frac{\left(-B_1\right)^n}{(n+m)!}L_n^{(m)}\left(\mu r^2 \right)
\sum_{j=0}^{\infty} \frac{\left( B_1\right)^j}{j!}H_{m+n+j}\left(B_1\right) 
\nonumber \\  \fl &+& 
\sum_{m=1}^{\infty}e^{-im\lambda(\vf-v)} \left(\mu r^2\left|{\cal L}_c\right|\right)^{m/2}
\sum_{n=0}^{\infty}\frac{\left(-B_1\right)^n}{(n+m)!}L_n^{(m)}\left(\mu r^2 \right)
 \sum_{k=0}^{\infty} \frac{\left( B_1\right)^k}{k!}H_{n+k}\left(B_1\right) 
\label{Psi-H}
\eeqn
Comparing (\ref{psiPsi}) and (\ref{Psi-H}) with (\ref{psiLag}) and using formula 5.12.1.3 from \cite{BrMar}
\be
\sum_{k=0}^{\infty} \frac{t^k}{k!}H_{n+k}(x) =\exp\left(2xt - t^2\right) H_n(x-t)
\ee
we arrive at the following expressions for coefficients $c_{n_r,m}$:
\be
c_{n_r,\lambda m} = \frac{\left(1-\eta^2\right)^{1/4} \left(-B_1\right)^{n_r}}{\sqrt{n_r! \left(n_r +|m|\right)!}}
e^{i\phi -\left|{\cal L}_c\right|/2} D_{n_r,m},
\ee
where $\phi=\sin(2w)\left|{\cal L}_c\right| \eta/2$ and
\be
D_{n_r,m} =\left\{
\begin{array}{cc}
\left(\eta e^{-i\lambda u}/2\right)^{m/2}H_{m+n_r}(0), & m \ge 0
\\
\left(\left|{\cal L}_c\right| e^{2i\lambda v}\right)^{|m|/2}H_{n_r}(0), & m < 0
\end{array}
\right.
\ee
We see that only even radial quantum numbers $n_r$ give nonzero contributions for $m<0$, whereas the parity of $n_r$
must coincide with the parity of $m$ for $m\ge 0$. Remember that
\be
H_{2k}(0)=  (-1)^k \frac{(2k)!}{k!}, \quad H_{2k+1}(0)=0.
\ee
Probabilities $\left|c_{n_r,\lambda m}\right|^2$ do not depend on phases $v$ and $u$. They satisfy the normalization condition,
which can be written as (we use here the notation $x=\eta/2$ and $y=\left|{\cal L}_c\right|$ to simplify the formula)
\be
\sum_{n=0}^{\infty}\sum_{m=1}^{\infty}\frac{(xy)^n}{n!(m+n)!}\left( x^m \left[H_{m+n}(0)\right]^2
+  y^m \left[H_{n}(0)\right]^2\right) = \frac{e^{y} -1}{\sqrt{1-4x}}.
\ee
This identity is not obvious at first glance, but it can be proven after some algebra. It is interesting that in spite of strong
`entanglement' between the `intrinsic' and `external' rotations (characterized by parameters $\eta$ and ${\cal L}_c$) in the
formulas for coefficients $c_{n_r,\lambda m}$,
these rotations are totally disentangled in formulas for the mean angular momentum, mean energy and their variances, as
was shown in the preceding sections.

\section{Isotropic charged oscillator and charged particle in a homogeneous magnetic field}
\label{sec-magosc}

The stationary Schr\"odinger equation for the two-dimensional isotropic oscillator in a homogeneous magnetic field $B$,
described by the Hamiltonian (in the circular gauge of the vector potential)
\be
\hat{H} = \frac1{2M}\hat{\bf p}^2 + \frac{M}{2}\tilde\omega^2 \hat{\bf r}^2 
- \omega_L \hat{L}_z,  \quad \omega_L =\frac{eB}{2Mc}, \quad \tilde\omega^2 = \omega^2 +\omega_L^2
\ee
was solved in polar coordinates by Fock \cite{Fock28}:
\be
\fl
\psi_{n_r m}(r,\vf) = \sqrt{\frac{\tilde\mu n_r!}{\pi \left(n_r +|m|\right)!}}
\left(\tilde\mu r^2\right)^{|m|/2}L_{n_r}^{(|m|)}\left(\tilde\mu r^2\right) \exp\left(-\,\frac{\tilde\mu}{2} r^2 +i m\vf\right),
\label{psiFock}
\ee
\be
E_{n_r m} =\hbar\tilde\omega\left(1 +|m| +2n_r\right) -\hbar\omega_L m. 
\label{Emag}
\ee
The solution (\ref{psiFock}) differs from (\ref{psiLag}) by the change $\mu \to \tilde\mu= M\tilde\omega/\hbar$.
The extremal Gaussian states with the fixed values of ${\cal L}_i$ and ${\cal L}_c$ are given by formula (\ref{G-polmin}) with
parameter $\mu$ replaced by $\tilde\mu$. Their mean energy equals
\be
{\cal E} = \hbar\tilde\omega\left(1 +\left|{\cal L}_i \right| + \left|{\cal L}_c \right| \right) 
-\hbar\omega_L \left({\cal L}_i+{\cal L}_c \right).
\ee
 The minimal energy is achieved for {\em co-rotating\/} packets with equal signs of ${\cal L}$ and $\omega_L$: 
\be
{\cal E}_{min}({\cal L}) = \hbar\tilde\omega + \hbar\left(\tilde\omega -  \left|\omega_L\right|\right) |{\cal L}| , \quad 
{\cal L}\omega_L \ge 0.
\ee
Such packets are superpositions of energy eigenstates with $n_r=0$ and coefficients (\ref{c0cor}).

In the  case of charged particle in a homogeneous magnetic field ($\omega=0$, $\tilde\omega=|\omega_L|$) 
we get (assuming $\omega_L >0$)
\be
{\cal E} = \hbar\omega_L\left[1 +\left|{\cal L}_i \right|(1-\lambda) + \left|{\cal L}_c \right|\left(1-\lambda_c\right) \right].
\ee
The absolute minimum
${\cal E}_{min} = \hbar\left|\omega_L\right| $ is achieved for all co-rotating packets with ${\cal L}\omega_L \ge 0$.
Of course this is explained by the well known infinite degeneracy of energy eigenstates in this special case.

The energy variance can be calculated in the same way as in section \ref{sec-fluct}. The result is  (for $\omega_L >0$)
\beqn
\sigma_E/(\hbar\omega_L)^2 &=&  
2\left(1-\lambda_c\right)\left(1-\lambda\right) \left|{\cal L}_c\right| \left[ \left|{\cal L}_i\right| 
 -  \sqrt{\left|{\cal L}_i\right| \left(1+\left|{\cal L}_i\right|\right)}\cos(2w) \right]
\nonumber \\ && 
+ 2\left|{\cal L}_c\right| \left(1-\lambda_c\right) + 4\left|{\cal L}_i\right| \left(1+\left|{\cal L}_i\right|\right)
\left(1-\lambda\right).
\label{sigEmag}
\eeqn
The variance equals zero for all packets whose directions of `internal' and `external' rotations coincide with the direction of the Larmor rotation:
$\lambda=\lambda_c=1$. The relative phase $w$ is important if only $\lambda=\lambda_c=-1$ (packets performing `co-rotation' in the
direction opposite to the Larmor rotation). For `anti-rotating' packets, the energy variance equals either $4\left|{\cal L}_c\right| $
(if $\lambda_c=-1$) or $8\left|{\cal L}_i\right| \left(1+\left|{\cal L}_i\right|\right)$ (if $\lambda=-1$).

Applying the propagator \cite{Kenn} (with $\mu_L =M\omega_L/\hbar$) 
\be
\fl
G({\bf r},{\bf r}';t) =\frac{\mu_L}{2\pi i \sin(\omega_L t)} \exp\left\{\frac{i\mu_L}{2}\left[\cot(\omega_L t)
\left({\bf r} -{\bf r}' \right)^2 -2\left(xy' -yx' \right)\right]\right\}
\label{Green-mag}
\ee
to the initial function (\ref{psi}), parametrized as in (\ref{compl})-(\ref{G}) with initial phases $u_0$
and $v_0$, one can see again that function $\psi(x,y;t)$ maintains its form, provided $u_0$ and $v_0$ are
replaced by time dependent phases 
\be
u(t)=u_0 +2\omega_L t(\lambda-1), \quad v(t)=v_0 +  \omega_L t\left(\lambda_c-1\right).
\label{uvt-mag}
\ee
Thus we see again that all packets with $\lambda=\lambda_c=1$ do not rotate at all, although they can possess arbitrary values of
`external' and `internal' angular momenta (but the same minimal possible total energy $\hbar\omega_L$).

\section{Free particle}
\label{sec-free}

In the case of free particle, the center of packet moves along a straight line, so that there is no `external' rotation.
Therefore it is sufficient to consider the special case of {\em homogeneous\/} Gaussian packets with the center fixed at point $(0,0)$.
There is no positive minimal energy for a free particle, therefore the results of this section cannot be obtained as a limit 
$\omega \to 0$ of the preceding sections. The qualitative difference is that the shape and size of free packets are inevitably deformed
during their evolution in time, whereas the minimum energy packets studied in the preceding sections maintain their shape and size.
For this reason we put hereafter $\mu=1$, using dimensional coefficients
$a$, $b$ and $c$.

The well known free particle propagator \cite{Kenn}
\be
G({\bf r},{\bf r}';t) =\left[2\pi i \hbar t/{m}\right]^{-1/2} \exp\left[\frac{i m}{2\hbar t}\left({\bf r}-{\bf r}'\right)^2 \right]
\label{Green-f}
\ee
transforms initial packet (\ref{psi}) to another Gaussian packet with time dependent coefficients
\be
a'=(a +i\tau D)/G(\tau), \quad c'=(c +i\tau D)/G(\tau), \quad b' = b/G(\tau),
\label{apr}
\ee
where
\be
\tau = 2\hbar t/m, \quad D= ac -b^2/4, \quad G(\tau) = 1 +i\tau(a+c) -\tau^2 D.
\ee

To diminish the number of parameters, let us consider the initial `most symmetrical' packet with nonzero mean angular momentum:
$\alpha=\gamma=\alpha_0$, $\chi_c=-\chi_a=\chi_0 >0$, $\beta_0>0$ and $\rho_0=0$. Then the conserved mean angular momentum is positive:
\be
{\cal L} =\frac{2\beta_0\chi_0}{\alpha_0^2 -\beta_0^2}.
\ee
The ellipses of constant probability densities have the semi-axes $a_{\pm}^2(0)=\nu/\left(\alpha_0 \mp\beta_0\right)$,
major semi-axis being inclined at the angle $-\pi/4$ with respect to $x$-axis.

However, the symmetry of coefficients is destroyed in the process of evolution, because real and imaginary parts of complex coefficients
(\ref{apr}) have the following form:
\be
\fl
\alpha(\tau) = \frac{\alpha_0}{F(\tau)}\left(1 +\tau^2 D_{+} -2\tau\chi_0\right), \quad
\gamma(\tau) =\frac{\alpha_0}{F(\tau)}\left(1 +\tau^2 D_{+} +2\tau\chi_0\right),
\label{algamt}
\ee
\be
\beta(\tau)=  \frac{\beta_0}{F(\tau)}\left(1 -\tau^2 D_{+} \right), \quad
\rho(\tau) = -\, \frac{\beta_0\alpha_0\tau}{F(\tau)}, \quad
\Delta(\tau) = \frac{\Delta(0)}{F(\tau)}
\label{beDel}
\ee
\be
\chi_{a,c}(\tau) = \left[\mp\chi_0\left(1 -\tau^2 D_{+} \right) - D_{-}\tau - D_{+}^2\tau^3\right]/F(\tau)
\ee
where
\be
F(\tau) =  1 + 2\tau^2 D_{-} + D_{+}^2\tau^4, \quad
D_{\pm}=  \frac14\left(\alpha_0^2 \pm 4\chi_0^2 \mp \beta_0^2\right).
\label{FD}
\ee
Note that functions $G(\tau)$ and $F(\tau)$ introduced above have no relation to the coefficients $G$ and $F$ used in section \ref{sec-Gauss}.

Contrary to the harmonic oscillator case, the free particle ellipse of constant relative probability not only rotates, but it changes its shape.
This is clearly seen from the formula describing the ellipse area $Y(t)$ or the probability density at the center of packet:
\be
Y(\tau)/Y(0)= |\psi(0,0;0)/\psi(0,0;\tau)|^2 = \sqrt{F(\tau)}.
\ee
While coefficient $D_{+}$ in equation (\ref{FD}) is always positive, coefficient $D_{-}$ is negative if 
$ 4\chi_0^2 > \alpha_0^2 +  \beta_0^2$. In such a case the packet {\em shrinks\/} initially, since 
$ |\psi(0,0;0)|^2 <|\psi(0,0;\tau)|^2 $ for small values of $\tau$. Such a behavior is not surprising for correlated ($\chi_0 \neq 0$)
Gaussian packets in one dimension, since parameter $\chi$, depending on its sign, is responsible for the effects of focusing or de-focusing 
(therefore correlated free Gaussian packets in one dimension were named `contractive states' in \cite{Yuen83,Stor94}). 
This is clearly seen from formula (\ref{j}), which shows that the initial probability current is directed
to/from the center if $\chi \neq 0$. A nontrivial  effect of quantum shrinking of packets with initial zero probability current
density was discovered in \cite{Schleich1} and generalized in \cite{DA1,DA2,Dahl04,Schleich2,Schleich3}. Note that packets studied
in this connection in two dimensions had the ring-shaped forms with zero probability density at the origin \cite{Schleich1,DA2,Dahl04}. We see that
the 2D Gaussian packets with the maximum of probability density at the origin can also shrink, but this time this happens
because of some competition between focusing and de-focusing in orthogonal directions ($\chi_a= -\chi_c$), and this
happens for sufficiently strong focusing only.

The minimal value 
\be
F_{min} = \frac{4\alpha_0^2\left(4\chi_0^2 -\beta_0^2\right)}
{\left(\alpha_0^2 +4\chi_0^2 -\beta_0^2\right)^2}
\ee
 is achieved for $\tau_{min}^2 = -D_{-}/D_{+}^2$. This value can be made as small as desired 
 for big enough values of parameter $\chi_0$ (so that the
packet can be concentrated in a very small region at $\tau=\tau_{min}$).
Note that this can happen even for $\beta_0=0$, i.e., for zero mean value of the angular momentum. This shows the
importance of parameter $\chi_0$, which determines the initial correlation coefficient between the coordinates and conjugated
momenta.

If $\beta_0 \neq 0$ (i.e., ${\cal L} \neq 0$), then some rotation of the packet is observed:
\be
\tan[2\theta(\tau)] =\frac{2\beta_0\left(1 -\tau^2 D_{+} \right)}{4\tau\chi_0\alpha_0}.
\label{theta-free}
\ee
The major axis becomes parallel to one of coordinate axes when $\tau^2 D_{+}=1$
(one can easily check that this happens after $\tau_{min}$ if $\tau_{min}$ exists). After that instant, angle $\theta$
changes its sign, going asymptotically to the same absolute value $\pi/4$ as at $t=0$. This means that the asymptotic direction of major axis
is perpendicular to the initial one.

The ellipse also changes its shape in the process of evolution: 
\be
a_{\pm}^2(\tau) =\frac{\nu}{\Delta_0}\left[\alpha_0 \left(1 +\tau^2 D_{+} \right) \pm
\sqrt{\beta_0^2\left(1 -\tau^2 D_{+} \right)^2 +4\alpha_0^2\chi_0^2\tau^2}\right].
\ee
For $\tau \ll 1$ we have
\[
a_{\pm}^2(\tau) \approx a_{\pm}^2(0)\left\{1 +\frac{\tau^2}{4|\beta_0|}\left[
|\beta_0| \left(\alpha_0 \mp |\beta_0|\right)^2 \pm 4\chi_0^2 \left(2\alpha_0 \mp |\beta_0|\right)\right]\right\}.
\]
Consequently, the major semi-axis always increases, whereas the minor one decreases for big enough values of $|\chi_0|$.
In particular, for sure it decreases for $4\chi_0^2 > \alpha_0^2 + \beta_0^2$.

To understand the evolution of the ellipse eccentricity we transform the corresponding expression in (\ref{apm}) as follows:
\[
\frac{2}{\vep^2} = 1 +\frac{\alpha +\gamma}{R} = 1 +\left[1- \frac{4\Delta}{(\alpha+\gamma)^2}\right]^{-1/2}.
\]
In turn, using (\ref{algamt}) and (\ref{beDel}) we have
\[
\frac{4\Delta}{(\alpha+\gamma)^2} = \frac{\Delta(0) F(\tau)}{\alpha_0^2\left(1 +\tau^2 D_{+} \right) ^2} =
\frac{\Delta(0) }{\alpha_0^2}\left[1-2\left(D_{+} -D_{-}\right)\left(D_{+}\tau + \frac1{\tau}\right)^{-2}\right].
\]
Consequently, the eccentricity attains the maximal value
\be
\vep_{max}^2 = \frac{2|\chi_0|}{|\chi_0| +\sqrt{D_{+}}} = \frac{4|\chi_0|}{2|\chi_0| + \sqrt{4\chi_0^2 + \alpha^2 -\beta^2}}
\label{vepmax}
\ee
exactly at the time instant $\tau_0$ when the ellipse axes become parallel to the coordinate ones: $\tau_0^2 D_{+}=1$.
At this instant we have
\be
a_{\pm}^2(\tau_0)=  \frac{2\alpha_0\, a_{\pm}^2(0)}{\alpha_0 \pm |\beta_0|}\left( 1 \pm \frac{|\chi_0|}{\sqrt{D_{+}}}\right),
\quad
F(\tau_0) = \frac{4\alpha_0^2}
{\alpha_0^2 +4\chi_0^2 -\beta_0^2}.
\ee
Obviously $F(\tau_0) > F_{min}$. Nonetheless $F(\tau_0)$ still can be smaller than unity for big values of $|\chi_0|$. 
In such a case, the area of ellipses at $\tau=\tau_0$ is smaller than the initial one.
Function $F(\tau)$ returns to the initial unity value when $\tau^2 = -2 D_{-}/D_{+}^2 = 2 \tau_{min}^2$.

\begin{figure}[htb]
\vspace{-0.5cm}
\begin{center}
\includegraphics[height=2.0truein,angle=0]{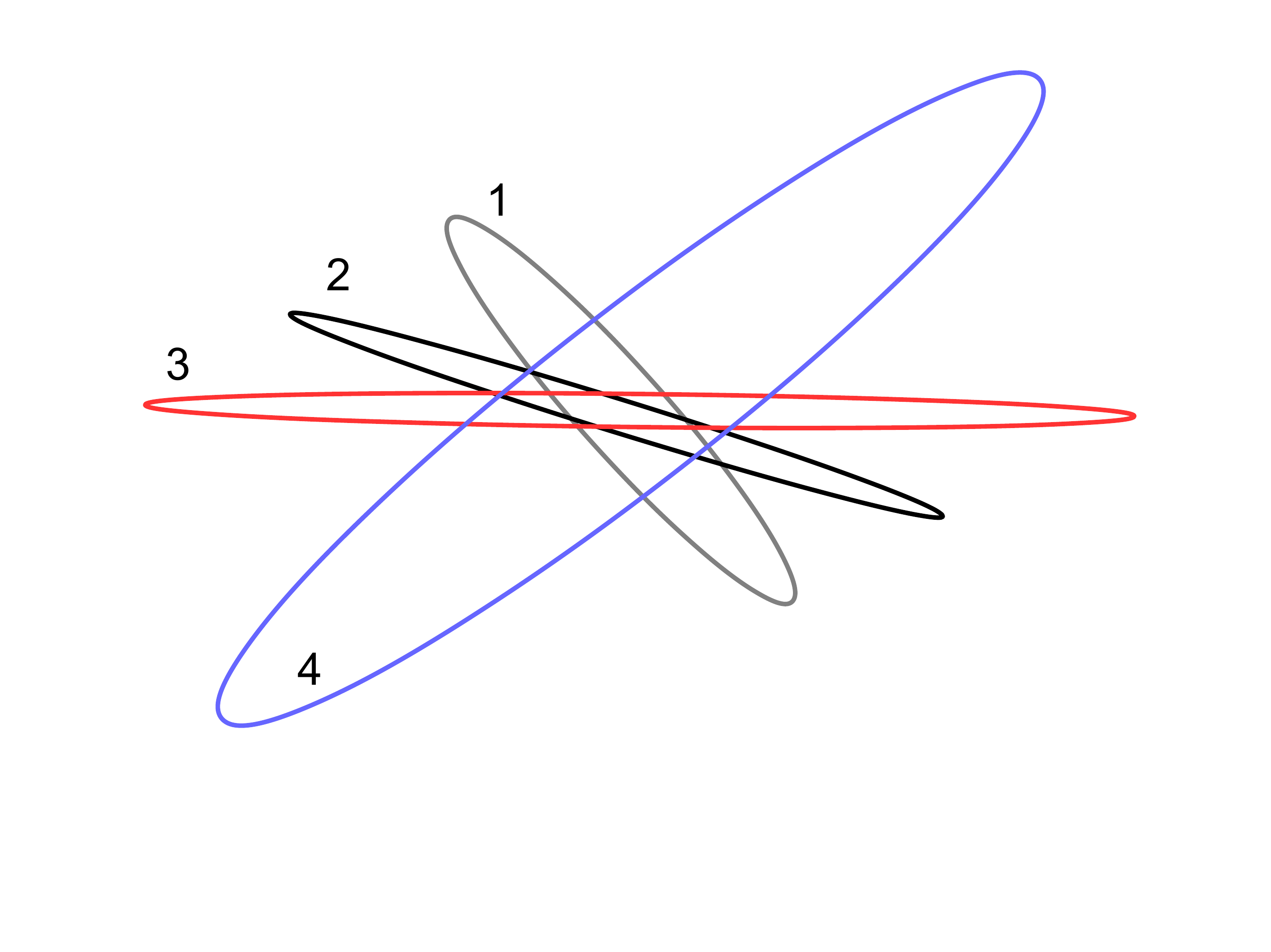}
\end{center}
\vspace{-1.5cm}
\caption{The orientation and shape of the ellipse of constant relative probability density of the free Gaussian packet
with big initial coordinate-momentum correlation coefficients at four specific time instants. 1 - the initial instant $\tau=0$.
2 - the instant $\tau_{min}$, when the probability density at the center attains the maximal value and the ellipse area is minimal.
3 - the instant $\tau_0$, when the eccentricity attains the maximal value and the ellipse is rotated by $45^{\circ}$ with respect
to the initial position. 4 - the asymptotical behavior: the eccentricity returns to the initial value, the ellipse is rotated by
almost $90^{\circ}$ and its area grows unlimitedly.
}
\label{fig-4ellip}
\end{figure}

When $\tau \to \infty$, then 
$ a_{\pm}(\tau) \approx a_{\pm}(0) \sqrt{D_{+}}\, \tau$ and the eccentricity returns to the initial value
$
\vep^2(0)=\vep^2(\infty) = {2|\beta_0|}/\left(\alpha_0 +|\beta_0|\right)$.
The probability density at the center decreases asymptotically as $|\psi(0,0;\tau)/\psi(0,0;0)|^2 \approx\left(D_{+}\tau^2\right)^{-1}$.
The change in time of the shape and orientation of the ellipses of constant relative probability density is shown
schematically in figure \ref{fig-4ellip}.

\section{Conclusion}
We have studied the properties of two-dimensional Gaussian packets with fixed mean values of angular momentum.
They depend on mutual directions of the independent `internal' and `external rotations' in the case of two-dimensional
harmonic isotropic oscillator. Moreover, the direction of the Larmor rotation is also important in the presence of an
additional homogeneous magnetic field. The states minimizing the total mean energy possess nonzero correlation
coefficients between coordinates and conjugated momenta. They also show a moderate squeezing of the quadrature
components. The distribution function over energy eigenstates can  exhibit some kind of `sub-Poissonian' statistics
for `co-rotating' packets satisfying certain conditions between the mean values of `internal' and `external' angular momenta.
In the case of free particle, packets with big enough initial coordinate--momentum  correlation coefficients
shrink initially. Only after some time they start to expand, rotating the directions of their major/minor  axes of constant probability 
density ellipses by $90$ degrees.
Since the shrinking effect can be very strong for the packets with big initial coordinate-momentum correlation coefficients,
such packets, perhaps, could find applications in the sensitive electron microscopy.
The time $\tau_{min}$ of the `maximal shrinking' can be adjusted to the necessary distance between the beam source 
and target by means of choosing the appropriate longitudinal velocity of the real three-dimensional beam along the
propagation axis.

\section*{Acknowledgment}

A partial support of the Brazilian agency  CNPq is acknowledged.

\appendix

\section{Proof of solution (\ref{g1}) }\label{ap1}

Using the relation $\Delta = g^2 -\eta^2$ one can represent the equation $\partial E_1/\partial \eta =0$ as 
 $\eta/\Delta^2 = g^2{\cal L}_i^2/\eta^3$, so that $\eta^2 = |{\cal L}_i| g\Delta$ (remember that $g>0$ and $\Delta >0$). 
Then it is easy to obtain the relations
\be
\eta^2 = \frac{|{\cal L}_i| g^3}{1+g|{\cal L}_i|}, \qquad
\Delta =  \frac{ g^2}{1+g|{\cal L}_i|}.
\label{A1}
\ee
The equation $\partial E_1/\partial g =0$ has the form
\be
1 + \frac1{\Delta} + \frac{{\cal L}_i^2 \Delta}{\eta^2} +2g^2\left(\frac{{\cal L}_i^2 }{\eta^2} -\,\frac1{\Delta^2} \right) =0.
\label{A2}
\ee
Putting expressions (\ref{A1}) into the left-hand side of (\ref{A2}) one arrives at the simple equation
$1-1/g^2 =0$, resulting in the solution  (\ref{g1}).

\end{document}